\documentclass[mathpazo]{cicp}
\usepackage{amsmath,amsfonts,amssymb,mathrsfs,xcolor,hyperref,booktabs}

\newcommand\maA{\mathcal{A}}
\newcommand\maB{\mathcal{B}}

\newcommand\bx{\boldsymbol{x}}
\newcommand\bz{\boldsymbol{z}}

\newcommand\bbf{\boldsymbol{f}}
\newcommand\bP{\boldsymbol{P}}

\newcommand\bk{\boldsymbol{\lambda}}

\begin{document}

\title{Numerical discretization of variational phase field model for phase transitions in ferroelectric thin films}

\author[Li R T et.~al.] {Ruotai Li\affil{1}, Qiang Du\affil{2}, Lei Zhang\affil{3} \comma\corrauth}
\address{\affilnum{1}\ Beijing International Center for Mathematical Research, Peking University, Beijing 100871, China\\
\affilnum{2}\ Department of Applied Physics and Applied Mathematics and Data Science Institute, Columbia University, New York, NY 10027, USA\\
\affilnum{3}\ Beijing International Center for Mathematical Research, Center for Quantitative Biology, Peking University, Beijing 100871, China}

\emails{{\tt liruotai@pku.edu.cn} (R. T.~Li), {\tt qd2125@columbia.edu} (Q.~Du), {\tt zhangl@math.pku.edu.cn} (L.~Zhang)}

\begin{abstract}
Phase field methods have been widely used to study phase transitions and polarization switching in ferroelectric thin films. In this paper, we develop  an efficient numerical scheme for the variational phase field model based on variational forms of the electrostatic energy and the relaxation dynamics of the polarization vector. The spatial discretization combines the Fourier spectral method with the finite difference method to handle three-dimensional mixed boundary conditions. It allows for an efficient semi-implicit discretization for the time integration of the relaxation dynamics. This method avoids explicitly solving the electrostatic equilibrium equation (a Poisson equation) and eliminates the use of associated Lagrange multipliers. We present several numerical examples including phase transitions and polarization switching processes to demonstrate the effectiveness of the proposed method.
\end{abstract}


\maketitle
\baselineskip 16pt
\section{Introduction}
In recent two decades, ferroelectric thin films have been given much attention both theoretically and experimentally \cite{Bratkovsky_prl_2000, Choi_Science_2004_enhance, Haeni_Nature_2004_room,  Jo_PRL_nonlinear_2009, Eliseev_prb_2012, Eliseev_prb_2013, Liu_Nature_intrinsic_2016}. These functional materials possess a spontaneous polarization that can be switched between energetically equivalent states in a single crystal by an electric field. Ferroelectric phase transitions and polarization switching depend on not only the stability of domain structures but also the electrostatic interactions or the external electric field \cite{Choudhury_APL_2008_effect}. 

Phase field methods have been successfully applied to study a wide range of physical problems, such as 
nucleation in solid-state phase transformations \cite{Zhang_npj_2016_recent}, coarsening process via epitaxial thin film model \cite{Qiao_2015_mathcompt},  phase transitions and domain structures in ferroelectric thin films \cite{Chen_JACS_2008_review}, etc. In the study of ferroelectric thin films, they characterize the detailed three-dimensional (3D) domain structures without any {\it a prior} assumptions with regard to the possible domain structures \cite{Li_Actamateria_2002_subs}. Phase field methods are able to predict not only the volume fractions of different orientation domains and the change of domain structures under the effect of applied external conditions, but also the temporal evolution of polarization during a ferroelectric phase transition \cite{Wang_Actamateria_2007_phase, Britson_Actamateria_2016_phase}.

In the phase field approach to study the ferroelectric phase transitions, the polarization vector is commonly used to describe a domain structure as the primary phase field variable. The electric potential, another variable, is then used to account for the electrostatic contributions in the phase field model \cite{Li_APL_2002_effect}. In the existing literature \cite{Wang_Actamateria_2007_phase,Britson_Actamateria_2016_phase, Li_APL_2002_effect, Li_APL_2006_temper}, the relation between the two variables is described by the electrostatic equilibrium equation (a Poisson equation), meaning that the electric potential is a function of the polarization vector and can be obtained by solving the electrostatic equilibrium equation when given a polarization distribution. While, the two variables are simply taken as independent ones when calculating the electrostatic driven force for the relaxation dynamics (electrostatic energy variation) of the polarization vector, ignoring their explicit relation given by the electrostatic equilibrium condition. Such treatment may simplify the calculation, but the effect of the electric field on phase transitions and polarization switching is underestimated. 

In a recent study \cite{fe_mod_2019}, new variational phase field formulations were proposed based on a hybrid representation in both real and Fourier variables in order to handle mixed electric boundary conditions (BCs) so that the coupling between the two phase field variables are directly incorporated. Indeed, the variational formulations allow a direct variational calculation of the phase field electrostatic energy and the driving force for the relaxation dynamics of the polarization vector. Furthermore, by utilizing the electrostatic equilibrium relation under the bound charge condition, the calculation of the electrostatic energy and its driving force can be done with respect to the polarization vector alone, thus simplifying the analytical derivations. Such variational forms can precisely and explicitly calculate the electrostatic energy and its corresponding driving force under different common-used electric BCs, e.g., the constant, open circuit, and tip-induced BCs.

In this paper, we develop an efficient numerical scheme for the variational phase field model to discretize the variational phase field formulations for effective 3D numerical simulations. It handles the 3D mixed BCs by combining the Fourier spectral method with the finite difference method for the spatial discretization. This in turn allows for the relaxation dynamics to be solved in a semi-implicit way. In particular, this  numerical scheme implements the calculation of the electric potential and the electrostatic driving force as matrix and vector multiplications at the discrete level that avoids explicitly solving the electrostatic equilibrium equation at each time step. The variational phase field model with its efficient numerical scheme can be used to investigate the effect of electrostatic interactions under different electric BCs on ferroelectric phase transitions and polarization switching as shown in numerical experiments.

The rest of the paper is structured as follows. The variational phase field formulations of electrostatic interactions are reviewed in  Sec. \ref{sec:model}. We present an efficient numerical scheme as the finite-dimensional discretization for the variational phase field formulations and its numerical implementation in Sec. \ref{sec:numer}. To show the effectiveness of the numerical scheme, we present 3D numerical simulations of the phase transitions and the polarization switching processes in the cubic thin film of lead titanate (PbTiO$_3$) in Sec. \ref{sec:result}. Some conclusions are given in Sec. \ref{sec:conclus}.

\section{Variational phase field formulations of electrostatic interactions}
\label{sec:model}
The ferroelectric domain structure in a thin film is often described by the spatial distribution of local polarization $\bP(\bx)=(P_1,P_2,P_3)$, with $\bx=(x,y,z)$ being the Cartesian coordinates in the 3D space. The temporal evolution of the polarization vector $\bP$ is given by the time dependent Ginzburg-Landau (TDGL) equations, which are the gradient dynamics of the total free energy $F$, as follows
\begin{equation}
\label{eq:tdgl}
\frac{\partial P_i(\boldsymbol{x},t)}{\partial t}=-\eta\frac{\delta F}{\delta P_i(\bx,t)},\quad i=1,2,3,
\end{equation}
where  $\eta$ is the kinetic coefficient related to the domain-wall mobility,  and $\delta F/\delta P_i(\bx,t)$ is the total driving force for  $P_i(\bx,t)$. We set the system in a thin film domain given by $\Omega=(-L/2,L/2)^2\times(0,h)$, where $L$ specifies the period along each of the directions parallel to the film and $h$ specifies the film thickness. The total free energy $F$ consists of three contributions: ferroelectric bulk free energy, domain wall energy, and electrostatic energy, which can be calculated by integrating their corresponding energy density function, i.e., $f_{bulk}$, $f_{wall}$ and $f_{ele}$ respectively, over the volume of the domain $\Omega$:  
\begin{equation}
\label{eq:energy}
F=\int_{\Omega} [f_{bulk}+f_{wall}+f_{ele}]\ d\bx.
\end{equation}
The mathematical expression for $f_{bulk}$, $f_{wall}$ and $f_{ele}$ can be found in \cite{Li_Actamateria_2002_subs, fe_mod_2019}, but as a quick review, the driven forces of bulk free energy and domain wall energy are expressed as
\begin{equation}
\begin{aligned}
\label{eq:landau}
b(P_i)=&\frac{\delta F_{bulk}}{\delta P_i }=2\alpha_1P_i+4\alpha_{11}P_i^3+2\alpha_{12}P_i\sum_{j\neq i}P_j^2 + 4\alpha_{112}P_i^3\sum_{j\neq i}P_j^2\\ 
&+2\alpha_{112}P_i\sum_{j\neq i}P_j^4+6\alpha_{111}P_i^5 +2\alpha_{123}P_i\prod_{j\neq i}P_j^2, \quad   i,j=1,2,3, 
\end{aligned}
\end{equation}
and
\begin{equation}
\begin{aligned}
\label{eq:wall}
w(P_i)=&\frac{\delta F_{wall}}{\delta P_i }=-G_{11}\frac{\partial^2 P_i}{\partial x_i^2}-(G_{44}+G'_{44})\sum_{j\neq i}\frac{\partial^2 P_i}{\partial x_j^2}\\
&+(G'_{44}-G_{12}-G_{44})\sum_{j\neq i}\frac{\partial^2P_j}{\partial x_i\partial x_j}, \qquad i,j=1,2,3,
\end{aligned}
\end{equation}
respectively, where $\alpha_1$, $\alpha_{11}$, $\alpha_{12}$, $\alpha_{111}$, $\alpha_{112}$, $\alpha_{123}$ are the Landau expansion coefficients, and $G_{11}$, $G_{12}$, $G_{44}$, $G'_{44}$ are the domain wall energy coefficients, with $(x_1,x_2,x_3)$ denoting the Cartesian coordinates $(x,y,z)$, respectively.
Natural Neumann-type boundary conditions of the form
 \begin{equation}
\left\{
\begin{aligned}
&
(G_{44}+G'_{44})\frac{\partial P_1}{\partial x_3}
+(G_{44}-G'_{44})\frac{\partial P_3}{\partial x_1} =0,\\
& (G_{44}+G'_{44})\frac{\partial P_2}{\partial x_3}
+(G_{44}-G'_{44})\frac{\partial P_3}{\partial x_2}=0, \\
&G_{11}\frac{\partial P_3}{\partial x_3}
+G_{12} \left(\frac{\partial P_1}{\partial x_1}+ \frac{\partial P_2}{\partial x_2}\right)=0,
\end{aligned} \right.  
\end{equation}
are also implied on the top ($x_3=h$) and bottom ($x_3=0$) surfaces.\par

In the existing literature \cite{Wang_Actamateria_2007_phase, Britson_Actamateria_2016_phase, Li_APL_2002_effect, Li_APL_2006_temper}, the most commonly used electrostatic energy density functions in the phase field models are $f_{ele}=\frac{1}{2}\nabla\phi(\bP)\cdot\bP$ and $f_{ele}=\frac{1}{2}(\nabla\phi(\bP)\cdot\bP+\epsilon|\nabla\phi(\bP)|^2)$, where $\epsilon$ is the dielectric permittivity and $\phi$ represents the electric potential with $-\nabla\phi$ being the electric field. Two energy density functions represent the electrostatic energy under the free charge condition and the bound charge condition, respectively \cite{fe_mod_2019}. 
According to these forms, if $\phi$ and $\bP$ are simply taken as independent variables, the electrostatic driven force $e_{old}(\bP)$, i.e., the electrostatic energy variation with respect to $\bP$, is given by
\begin{equation}
\label{eq:oldphi}
e_{old}(\bP)=\frac{1}{2}\nabla\phi(\bP) \ .
\end{equation}
In fact, such electrostatic driven force is inaccurate because two phase field variables $\phi$ and $\bP$ are related and satisfy the electrostatic equilibrium condition ($\Delta\phi=\nabla\cdot\bP$) \cite{Wang_Actamateria_2007_phase,Britson_Actamateria_2016_phase, Li_APL_2002_effect, Li_APL_2006_temper}. In a recent work \cite{fe_mod_2019}, Du et al. proposed new variational phase field formulations of the electrostatic energy, which avoid this ambiguity in previous studies. By the variational formulations in \cite{fe_mod_2019}, the electrostatic driven force $e(\bP)$ subject to various electric BCs is written as 

\begin{equation}
\label{eq:tab1}
e(\bP)=
\left\{
\begin{array}{ll}
\nabla\phi(\bP)-\frac{1}{2}\nabla\phi_2\, , \quad   &\mbox{Dirichlet BC, e.g., constant BC,  tip BC, }\\
 \nabla\phi(\bP) \, , &\mbox{Neumann BC, e.g., open circuit BC.}
 \end{array}
\right.
\end{equation}
Here $\phi_2$ is an auxiliary potential. 
It is clear to show that the electrostatic driven force $e(\bP)$ based on the variational formulations is almost twice as $e_{old}(\bP)$ . 
The effect of $e(\bP)$ and $e_{old}(\bP)$ on the ferroelectric phase transitions will be evaluated by numerical simulations in Sec. \ref{sec:result}. 

Due to the periodicity of  $\phi$ and $\bP$ in the $x $-$y$ plane, we use $\hat{(\cdot)}$ to denote the 2D Fourier series expansion with $\bk=(\lambda_1,\lambda_2)$ being the variables in the Fourier (frequency) space. Under Dirichlet BC with $c_2$ and $c_1$ denoting the top and bottom boundary data of the film respectively,  $\phi_2$ can be recovered from its Fourier representation given by
\begin{equation}
\hat{\phi}_2(\lambda_1,\lambda_2,z)=\frac{\hat{c}_2-\hat{c}_1e^{-|\lambda|h}}{M(h)}e^{|\lambda| z}+\frac{\hat{c}_1e^{|\lambda|h}-\hat{c}_2}{M(h)}e^{-|\lambda| z},
\end{equation}
where $|\lambda|=\sqrt{\lambda_1^2+\lambda_2^2}$, $M(h)=e^{|\lambda| h}-e^{-|\lambda| h}$, and $\hat{c}_1$ and $\hat{c}_2$ are the Fourier representations of  $c_1$ and $c_2$ respectively.
The function $\phi$ can be also recovered from its Fourier representation given by
\begin{equation}
\label{eq:solve1}
\hat{\phi} (\lambda_1,\lambda_2, z)=C_1(\lambda_1,\lambda_2)e^{|\lambda| z}+C_2(\lambda_1,\lambda_2)e^{-|\lambda| z}+g(\lambda_1,\lambda_2, z),
\end{equation}
where the function $g=g(\lambda_1,\lambda_2,z)$ is defined by
\begin{equation}
\label{eq:gz}
g(\lambda_1,\lambda_2,z) 
= \frac{1}{2|{\lambda}|\epsilon}\int_0^{z} [ (e^{|\lambda|(z-s)}-e^{-|\lambda|(z-s)})  f(\lambda_1,\lambda_2,s)  ]\ ds,
\end{equation}
and $f=f(\lambda_1,\lambda_2, z)$ denotes the 2D Fourier representation of divergence of the polarization vector $\bP$, i.e., 
$f(\lambda_1,\lambda_2, z)=I\lambda_1\hat{P}_1+I\lambda_2\hat{P}_2+\frac{\partial\hat{P}_3}{\partial z}$ with $I=\sqrt{-1}$. 
The values of the coefficients $\boldsymbol{C}=(C_1,C_2)^\intercal$ are decided by the various BCs. For the Dirichlet BC and open circuit BC,  we have respectively 
\begin{equation}
\label{eq:Coe2}
 \boldsymbol{C}(\lambda_1,\lambda_2)=\frac{1}{ M(h) }\begin{pmatrix}
 -\hat{c}_1(\lambda_1,\lambda_2)e^{-|\lambda|h}+\hat{c}_2(\lambda_1,\lambda_2)-g(\lambda_1,\lambda_2,h) \\
 \hat{c}_1(\lambda_1,\lambda_2)e^{|\lambda|h}-\hat{c}_2(\lambda_1,\lambda_2)+g(\lambda_1,\lambda_2,h)
\end{pmatrix}
\end{equation}
and
\begin{equation}
\label{eq:Coe3}
\boldsymbol{C}(\lambda_1,\lambda_2)=\frac{1}{ |\lambda| M(h) } \begin{pmatrix}
 -\hat{P}_3(\lambda_1,\lambda_2,0)e^{-|\lambda|h}+\hat{P}_3(\lambda_1,\lambda_2,h)- g_3(\lambda_1,\lambda_2,h) \\
 -\hat{P}_3(\lambda_1,\lambda_2,0)e^{|\lambda|h}+\hat{P}_3(\lambda_1,\lambda_2,h)-g_3(\lambda_1,\lambda_2,h)
\end{pmatrix},
\end{equation} 
where 
\begin{equation}
\label{eq:gzd} 
g_3(\lambda_1,\lambda_2,h)=\frac{1}{2\epsilon}\int_0^{h}
[(e^{|\lambda|(z-s)}+e^{-|\lambda|(z-s)})f(\lambda_1,\lambda_2,s)] \ ds
\end{equation} 
denotes the value of the partial derivative of $g(\lambda_1,\lambda_2,z)$ to the third variable $z$ when $z$ equals $h$. 

With the explicit expression of $\hat{\phi}$, its gradient $\nabla\phi$ can be easily obtained from its Fourier expansion (or the discrete inverse Fourier transform on the lattice points), e.g., 
\begin{equation}
\label{eq:invphi1}
\left( \begin{array}{c}
\dfrac{\partial\phi}{\partial x}\\
\dfrac{\partial\phi}{\partial y}
\end{array}\right)
={L^2}\sum_{\lambda_1,\lambda_2=-\infty}^{\infty}I
\left(\begin{array}{c} \lambda_1\\
\lambda_2
\end{array}\right)
\hat{\phi}(\lambda_1,\lambda_2,z)e^{I\frac{2\pi\lambda_1x}{L}}e^{I\frac{2\pi\lambda_2y}{L}}
\end{equation}
and
\begin{equation}
\label{eq:invphi2}
\frac{\partial\phi}{\partial z}=L^2\sum_{\lambda_1,\lambda_2=-\infty}^{\infty}[|\lambda|(C_1e^{|{\lambda}| z}-C_2e^{-|\lambda| z})+g_3(z)]e^{I\frac{2\pi\lambda_1x}{L}}e^{I\frac{2\pi\lambda_2y}{L}}.
\end{equation}
In this way, the gradient of $\phi_2$ can also be computed, e.g., through Eq. \eqref{eq:invphi1}, and
\begin{equation}
\label{eq:invphi3}
\frac{\partial\phi_2}{\partial z}=L^2\sum_{\lambda_1,\lambda_2=-\infty}^{\infty}|\lambda|(\frac{\hat{c}_2-\hat{c}_1e^{-|\lambda|h}}{M(h)}e^{|\lambda| z}-\frac{\hat{c}_1e^{|\lambda|h}-\hat{c}_2}{M(h)}e^{-|\lambda| z})e^{I\frac{2\pi\lambda_1x}{L}}e^{I\frac{2\pi\lambda_2y}{L}}.
\end{equation}

Please note that the expressions of $\phi$ and $\phi_2$ shown above are valid for the case $(\lambda_1,\lambda_2)\neq(0,0)$. For the very special case  
$(\lambda_1,\lambda_2)=(0,0)$, it is easy to get the electric potential $\phi$ as well as the electrostatic driven force. The detailed derivations for the electric potential $\phi$ and the electrostatic driven force under different BCs are referred to \cite{fe_mod_2019}.

\section{Numerical scheme for variational phase field model}
\label{sec:numer}
In this section, we present the numerical discretization for the variational phase field model to effectively simulate phase transitions in ferroelectric thin films. 
A natural approach is to make use of the periodicity of $\phi$ and $\bP$ in the $x$-$y$ plane in Eq. \eqref{eq:tdgl}, as in the previous study  \cite{Chen_ComPhyComm_1998_apply} where the system is under the periodic BC and a semi-implicit discretization by the Fourier spectral approximation is used. 
Since we deal with the mixed BCs (different electric BCs in the $z$ direction and with periodicity in the $x$ and $y$ directions), we combine the Fourier spectral method in  the $x$-$y$ plane with the finite difference discretization in the $z$ direction to efficiently calculate the total driving force in Eq. \ref{eq:tdgl} and solve it in a semi-implicit way.

Following the variational phase field formulations together with the relations given in Eqs. \eqref{eq:invphi1}-\eqref{eq:invphi3}, we have a coupled ODE system that can be  written as
\begin{equation}
\label{eq:f1}
-\frac{1}{\eta}\frac{\partial}{\partial t}\hat{P_i}(\bk,z,t)=\hat{b}(P_i)_{\bk}+\hat{w}(P_i)_{\bk}+\hat{e}(\bP)_i, \qquad  i=1,2,3,
\end{equation}
where $\bk=(\lambda_1,\lambda_2)$ is the same as above. $\hat{b}(P_i)_{\bk}$, $\hat{w}(P_i)_{\bk}$, and $\hat{e}(\bP)_i$ represent 2D Fourier series expansions of $b(P_i)$, $w(P_i)$, and the $i$-th component of $e(\bP)$ respectively, for $i=1,2,3.$ 

For a given $\bP$ at each time step, the nonlinear term $\hat{b}(P_i)_{\bk}$ can be obtained directly by applying 2D pseudo-spectral Fourier series expansion to $b(P_i)$,  $i=1,2,3$. To calculate $\hat{w}(P_i)_{\bk}$, we adopt the second-order central difference scheme on a uniform grid in the $z$ direction to numerically approximate the partial derivative of $P_i$ to the third variable $z$. We assume that the number of grid points along the $z$ direction is $N_z$.  For given $\lambda_1$ and $\lambda_2$,  $\hat{w}(P_i)_{\bk}$, $N_z\times 1$ vector, can be written as 
\begin{equation}
\label{eq:wp}
\hat{w}(P_i)_{\bk}=
\left\{
\begin{array}{lll}
G_{11}\lambda_i^2\hat{P}_i+G_1(\lambda_2^2\hat{P}_i+\maB\hat{P}_i)+G_2(-\lambda_1\lambda_2\hat{P}_2+I\lambda_i\maA\hat{P}_3),\  i=1,\\  
G_{11}\lambda_i^2\hat{P}_i+G_1(\lambda_1^2\hat{P}_i+\maB\hat{P}_i)+G_2(-\lambda_1\lambda_2\hat{P}_1+I\lambda_i\maA\hat{P}_3) ,\ i=2,\\
G_{11}\maB\hat{P}_i+G_1(\lambda_1^2+\lambda_2^2)\hat{P}_i+G_2(I\lambda_1\maA\hat{P}_1+I\lambda_2\maA\hat{P}_2),\  i=3,
\end{array}\right. 
\end{equation}
where $I=\sqrt{-1}$ , $G_1=G_{44}+G'_{44}$ and $G_2=G_{44}-G_{12}-G'_{44}$. $\maB$ and $\maA$ are the $N_z\times N_z$ sparse matrices respectively, which can be written as 
\begin{equation*}
\maB=\frac{1}{(\Delta z)^2}\begin{pmatrix}
2 &-1  &0 &\cdots &0 &-1\\
0  &-1  &2  &1 &0  &\cdots\\
 &\ddots &\ddots  &\ddots &\ddots  \\
\cdots &0 &0 &-1 &2 &1\\
  -1 &0 &\cdots  &0 &-1 &2
\end{pmatrix} 
\end{equation*}
and
\begin{equation*}
\maA=\frac{1}{2\Delta z}\begin{pmatrix}
-1 &1 \quad &\cdots \\
-1  &0 &1 & \cdots\\
\qquad &\ddots \\
\cdots &-1 &0 &1\\
   &\cdots   &-1 &1
\end{pmatrix},
\end{equation*}
with $\Delta z=\frac{h}{N_z-1}$ being the unit grid length in the $z$ direction.\par

Lastly, to calculate the electrostatic driven force $\hat{e}(\bP)$ under different electric BC given by Eq. \eqref{eq:tab1}, we need to numerically compute the electric potential $\phi$ and its gradient $\nabla\phi$. From Eq. \eqref{eq:invphi1} and Eq. \eqref{eq:invphi2}, we notice that the computational complexity mainly comes from the calculation of $g(\bk,z)$ and $g_3(\bk,z)$ given in Eq. \eqref{eq:gz} and Eq. \eqref{eq:gzd}, respectively. It is important to highlight that $f=f(\bk,z)$ is solely computed from the polarization field $\bP$, so are the functions $g$, $g_3$ and $\hat{\phi}$. Thus, the numerical method for the calculation of integrals related to $g$ and $g_3$ is largely dependent on the way we discretize the equations along the $z$ direction, or vice versa. That is to say, the selection of the numerical quadrature method and the discretization used to solve equations on the same grid should be carried out together.\par

A simple choice is to adopt a finite difference approximation in the $z$ direction on a uniform grid, which allows us to conveniently apply the composite Simpson's rule based on the same grid points without further interpolations. For illustration, assuming the $\{z_j\}_1^{N_z}$ are evenly spaced grid points among interval $[0,h]$ with $z_1=0$ and $z_{N_z}=h$. For fixed $\lambda_1$ and $\lambda_2$,  $ \bbf=(f(\lambda_1,\lambda_2,z_1),\ldots, f(\lambda_1,\lambda_2,z_{N_z}))^\intercal$ becomes an $N_z\times 1$ vector, so are the polarization field $\hat{P_i}(\lambda_1,\lambda_2,\bz), i=1,2,3$, with $\bz=(z_1,z_2,\ldots,z_{N_z})^\intercal$.  We can easily get $\bbf(\bk,\bz)=I\lambda_1\hat{P}_1(\lambda_1,\lambda_2,\bz)+I\lambda_2\hat{P}_2(\lambda_1,\lambda_2,\bz)+\maA\hat{P}_3(\lambda_1,\lambda_2,\bz)$, where $\maA$ is the same matrix as above. Similarly, the calculation of $g(\bk,z_j)$ and $g_3(\bk,z_j)$ in this scheme becomes 
\begin{equation}
\label{eq:gz1}
g(\bk,z_j)=\frac{\Delta z}{12|\lambda|\epsilon}(e^{|\lambda|z_j}\mathcal{E}_j^{-}\mathcal{K}_j \bbf_j-e^{-|\lambda|z_j}\mathcal{E}_j^{+}\mathcal{K}_i\boldsymbol{f}_j), \quad j=1,2,\ldots,N_z,
\end{equation}
and 
\begin{equation}
\label{eq:gzd1}
g_3(\bk,z_j)=\frac{\Delta z}{12\epsilon}(e^{|\lambda|z_j}\mathcal{E}_j^{-}\mathcal{K}_j \bbf_j+e^{-|\lambda|z_j}\mathcal{E}_j^{+}\mathcal{K}_j\boldsymbol{f}_j), \quad j=1,2,\ldots,N_z,
\end{equation}
respectively, where 
$$\mathcal{E}_j^{-}=(e^{-|\lambda| z_1},e^{-|\lambda|(z_1+z_2)/2},e^{-|\lambda| z_2},\ldots,e^{-|\lambda| z_{j-1}},e^{-|\lambda|(z_{j-1}+z_j)/2},e^{-|\lambda| z_j})$$ 
is a $1\times (2j-1)$ vector, and
$$\mathcal{E}_j^{+}=(e^{|\lambda| z_1},e^{|\lambda|(z_1+z_2)/2},e^{|\lambda| z_2},\ldots,e^{|\lambda| z_{j-1}},e^{|\lambda|(z_{j-1}+z_j)/2},e^{|\lambda| z_j})$$ 
is also a $1\times (2j-1)$ vector. Meanwhile, 
$$\boldsymbol{f}_j=(f(\bk,z_1),\frac{f(\bk,z_1)+f(\bk,z_2)}{2},f(\bk,z_2),\ldots,\frac{f(\bk,z_{j-1})+f(\bk,z_j)}{2}, f(\bk,z_j))^\intercal$$
 is a $(2j-1)\times 1$ vector, and $\mathcal{K}_j$ is a $(2j-1)\times (2j-1)$ diagonal matrix. More specifically, $\mathcal{K}_1=1$, $\mathcal{K}_2=diag(1,4,1)$, and $\mathcal{K}_j=diag(1,4,2,4,2,\ldots,4,2,4,1)$ for $3\leq j\leq N_z$, where $diag(\cdot)$ represents the main diagonal elements. The calculation of $g(\bk,z_j)$ and $g_3(\bk,z_j)$ for $j=1,2,\ldots,N_z$ at each time step becomes the calculation of $\bbf(\bk,\bz)$ in combination with the computation of Eq. \eqref{eq:gz1} and Eq. \eqref{eq:gzd1}, respectively. We note that at each time step only $\bbf(\bk,\bz)$ needs to be updated, and all the calculations actually turn into a number of matrix and vector multiplications that greatly reduce the computational complexity.
 
 With $g(\bk,z)$ and $g_3(\bk,z)$, the value of $\hat{e}(\bP)$ can be easily obtained from Eq. \eqref{eq:invphi1} and Eq. \eqref{eq:invphi2}. Thus, the semi-implicit hybrid Fourier and finite difference scheme for Eq. \eqref{eq:f1} becomes
\begin{equation}
\label{eq:f2}
\hat{P}_i^{n+1}+\eta\Delta t \hat{w}^{*}(P_i^{n+1})_{\bk}=\hat{P}_i^n-\eta\Delta t[\hat{b}(P_i^n)_{\bk}+\hat{e}(\bP^n)_i+\hat{w}^{**}(P_i^{n})_{\bk}], \quad i=1,2,3,
\end{equation}
where 
\begin{equation*}
\hat{w}^{*}(P_i^{n+1})_{\bk}=
\left\{ \begin{array}{lll}
G_{11}\lambda_i^2\hat{P}_i^{n+1}+G_1(\lambda_2^2\hat{P}_i^{n+1}+\maB\hat{P}_i^{n+1}),\quad i=1,\\
G_{11}\lambda_i^2\hat{P}_i^{n+1}+G_1(\lambda_1^2\hat{P}_i^{n+1}+\maB\hat{P}_i^{n+1}),\quad i=2,\\
G_{11}\maB\hat{P}_i^{n+1}+G_1(\lambda_2^2\hat{P}_i^{n+1}+\lambda_1^2\hat{P}_i^{n+1}),\quad i=3,
\end{array}\right.
\end{equation*}
and 
\begin{equation*}
\hat{w}^{**}(P_i^{n})_{\bk}=
\left\{\begin{array}{lll}
G_2(-\lambda_1\lambda_2\hat{P}_2^n+I\lambda_i\maA\hat{P}_3^n),\quad  i=1,\\  
G_2(-\lambda_1\lambda_2\hat{P}_1^n+I\lambda_i\maA\hat{P}_3^n) ,\quad i=2,\\
G_2(I\lambda_1\maA\hat{P}_1^n+I\lambda_2\maA\hat{P}_2^n),\quad  i=3.
\end{array}\right.
\end{equation*}
Here, $-\frac{1}{2}N_x+1\leq\lambda_1,\lambda_2\leq\frac{1}{2}N_x$ with $N_x$ being the maximum frequency in Fourier space, $G_1$ and $G_2$ are the same as before, and $\Delta t$ represents the time step size. 
The semi-implicit scheme in Eq. \eqref{eq:f2} allows a larger time step compared to the explicit Euler scheme. Nevertheless, the order of accuracy is only first order in $\Delta t$. Higher-order accuracy in time can be achieved by
using various high-order semi-implicit schemes for time discretization \cite{du2019phase}. For instance, one can apply a two-step backward difference for $(\partial/\partial t)\hat{P_i} (i=1,2,3)$ and two-step Adams-Bashforth for explicit treatment of nonlinear term in Eq. \eqref{eq:f1}, which leads to the following scheme:
\begin{equation}
\label{eq:f3}
\beta(\hat{P}_i^{n+1})=4\hat{P}_i^n-\hat{P}_i^{n-1}-\eta\Delta t[\hat{b}^{*}(P_i^n)_{\bk}+\hat{e}^{*}(\bP^n)_i+\hat{w}^{***}(P_i^{n})_{\bk}], \quad i=1,2,3,
\end{equation}
where $ \beta(\hat{P}_i^{n+1})=3\hat{P}_i^{n+1}+2\eta\Delta t\hat{w}^{*}(P_i^{n+1})_{\bk}$, $\hat{b}^{*}(P_i^n)_{\bk}=3\hat{b}(P_i^n)_{\bk}-\hat{b}(P_i^{n-1})_{\bk}$, $\hat{e}^{*}(\bP^n)_i=3\hat{e}(\bP^n)_i-\hat{e}(\bP^{n-1})_i$, and $\hat{w}^{***}(P_i^n)_{\bk}=3\hat{w}^{**}(P_i^n)_{\bk}-\hat{w}^{**}(P_i^{n-1})_{\bk}$.  In practice, we could first use Eq. \eqref{eq:f2} to compute $\bP^1$, and then adopt the scheme in Eq. \eqref{eq:f3}. Similarly, third-order semi-implicit schemes and even more sophisticated schemes could be designed by using the same idea. 

\section{Numerical examples}
\label{sec:result}
We now present several numerical examples to simulate the phase transitions and the polarization switching processes of the PbTiO$_3$ thin film by applying the the numerical discretization of the variational phase field formulations. 3D computational simulations are performed on the domain of size: $64\Delta x\times64\Delta x\times64\Delta x$, with a uniform grid spacing $\Delta x=1.0$ nm in all three coordinate directions. The coefficients of the bulk free driven force are taken exactly from \cite{Li_Actamateria_2002_subs}, and the dielectric permittivity is taken as $\epsilon=8.85\times10^{-10}Fm^{-1}$. The domain wall energy coefficients are taken isotropically and chosen to be $G_{11}/G_{110}=\frac{1}{2\Delta x}$, $G_{12}/G_{110}=0$, and $G_{44}/G_{110}=G'_{44}/G_{110}=\frac{1}{4\Delta x}$, where $G_{110}$ is related to the magnitude of grid spacing $\Delta x$ via $\Delta x=\sqrt{G_{110}/\alpha_0}$ and $\alpha_0=1.7252\times10^8C^{-2}m^2N$. \par

\subsection{Phase transitions of the PbTiO$_3$ thin films}
To represent the equivalent polarization magnitude and the corresponding polarization direction, i.e., $\bP=(0,0,-1)$ and $\bP=(0,0,1)$, blue and red colors are used respectively in the figures. The gradual change from blue color to red color represents the change of local polarization magnitude and direction from $-1$ to $1$ along the Z-axis, or vice versa. 

\begin{figure}[htbp]
\centering
\includegraphics[scale=0.95]{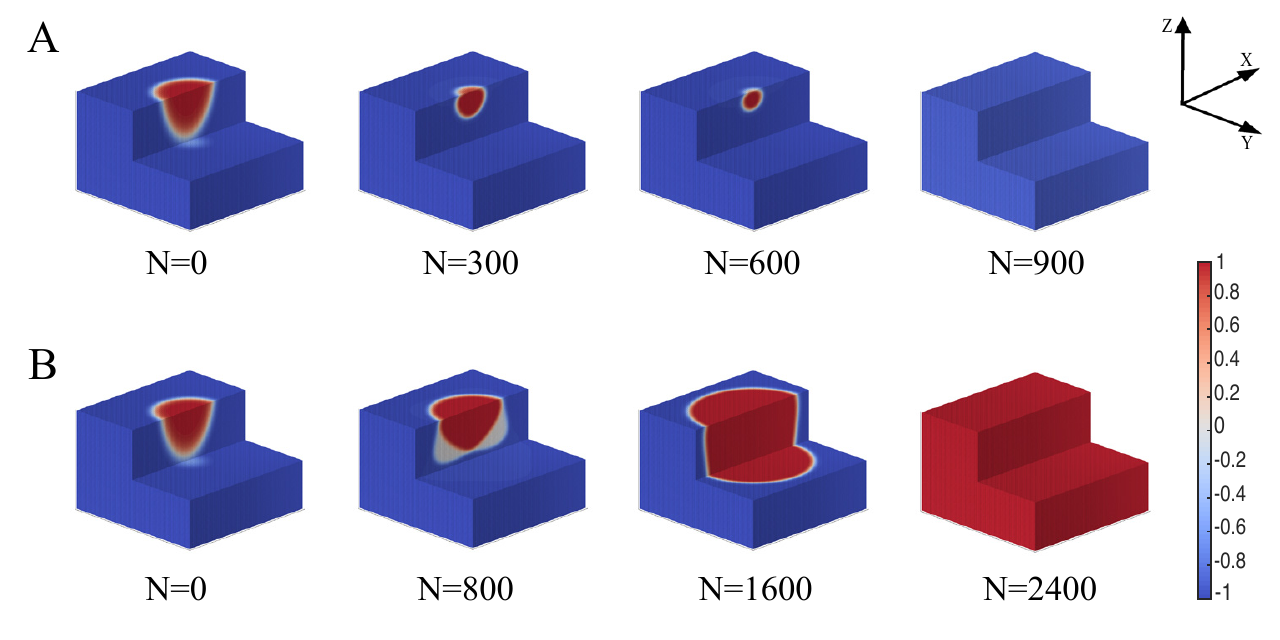}
\caption{\textbf{Ferroelectric phase transitions driven by two different electrostatic forces under the tip-induced BC.} (A): the electrostatic force $e_{old}(\bP)$ by Eq. \ref{eq:oldphi}; (B): the electrostatic force $e(\bP)$ by Eq. \ref{eq:tab1}.}
\label{fig:1}
\end{figure}
Fig. \ref{fig:1} shows the ferroelectric phase transitions from a tip-like domain configuration to two different equilibrium states under the tip-induced BC. 
The initial tip-induced electric potentials in both cases are the same, which are negative on the top surface and zero on the bottom surface of the film.  
In the case  the electrostatic force $e_{old}(\bP)$ given by Eq. \eqref{eq:oldphi}  is applied, the result shows that the electric field does not have a strong enough effect to induce the whole domain polarization switching in Fig. \ref{fig:1}A. However,  when the electrostatic force $e(\bP)$ given by Eq. \eqref{eq:tab1} is applied, it can induce a polarization switching under the same initial condition in Fig. \ref{fig:1}B.

\begin{figure}[htbp]
\centering
\includegraphics[scale=0.95]{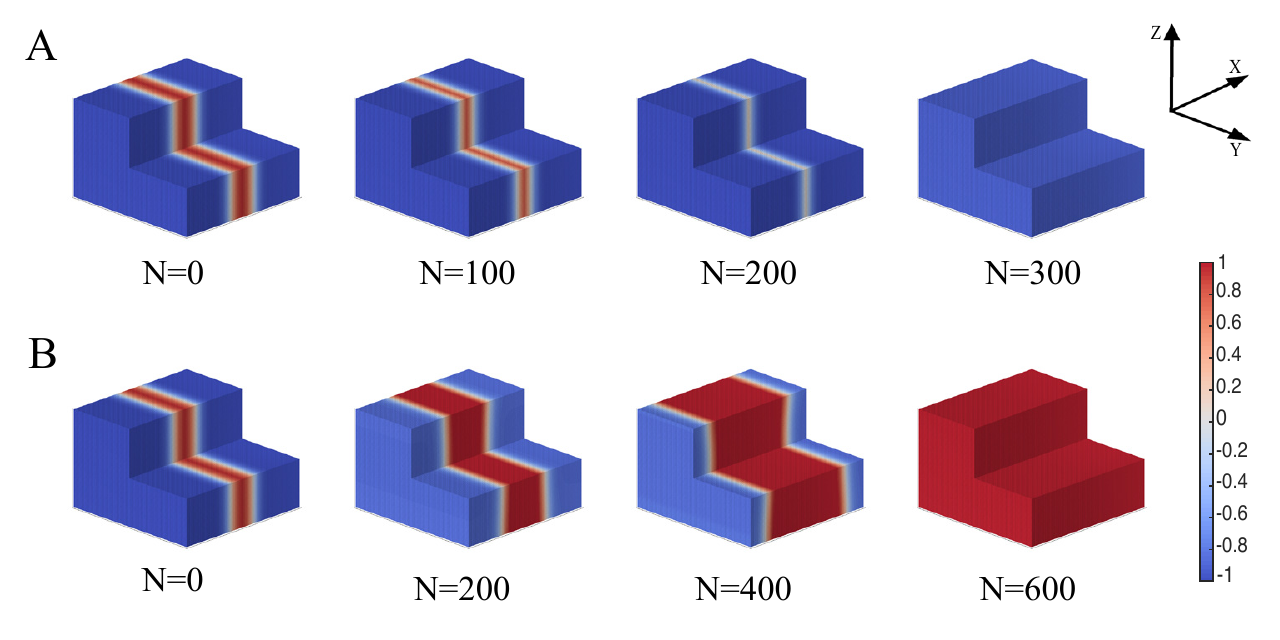}
\caption{\textbf{Ferroelectric phase transitions driven by two different electrostatic forces under the constant BC.} (A): the electrostatic force $e_{old}(\bP)$; (B): the electrostatic force $e(\bP)$.}
\label{fig:2}
\end{figure}

Fig. \ref{fig:2} shows the ferroelectric phase transitions starting from a ribbon-like domain configuration under the constant BC by applying the electrostatic driven force $e_{old}(\bP)$ and $e(\bP)$, respectively. Both electric potentials under constant BC are negative on the top surface and positive on the bottom surface of the film. 
The numerical results show the phenomenon is similar to that in Fig. \ref{fig:1}. We can observe that, compared to the accurate electrostatic force $e(\bP)$, the electrostatic force $e_{old}(\bP)$ obtained by simply treating the $\phi$ and $\bP$ as two independent variables, which is wrongly derived, indeed underestimates the electric effect on the polarization switching. 

To test the spatial accuracy of the proposed numerical scheme, we start from the same domain structure (Fig. \ref{fig:1}B) and compute the steady state with different mesh sizes. The steady state obtained by the mesh size of 256$\times$256$\times$256 is taken as the benchmark approximation of the exact solution. As the result shown in Table \ref{tab1}, this numerical scheme enjoys the second-order accuracy in space. Although the Fourier spectral discretization provides excellent spatial accuracy, the Eq. \eqref{eq:f2} is discretized by using a second-order central difference approximation along the $z$ direction, and thus the total order of spatial accuracy is limited by the finite difference approximation. If one wants to achieve a better spatial accuracy, we may use spectral approximations, e.g., Fourier sine and cosine transforms, or other orthogonal polynomials, e.g., Chebyshev polynomials and Legendre polynomials, along the $z$ direction. In this case, the discretization of equations along the $z$ direction also needs to be changed accordingly. 

\begin{table}[htbp]
\centering
\label{tab1}
\begin{tabular}{ccc}
\toprule
grid & $L^{\infty}$ &order\\
\midrule
16$\times$16$\times$16  & 4.0479e-2    & $-$   \\
32$\times$32$\times$32  &1.0056e-2     & 2.01   \\
64$\times$64$\times$64  &2.4624e-3     &2.03   \\
128$\times$128$\times$128  &6.1989e-4    &1.99  \\
\bottomrule
\end{tabular}
\caption{Order of spatial accuracy for the numerical scheme with grid refinement}
\end{table}

We also compare the maximum time step size for the proposed numerical scheme and the explicit scheme. 
The maximum time step size is estimated from the numerical simulations with the mesh size: 64$\times$64$\times$64. The explicit scheme means that the second-order partial derivative in Eq.\eqref{eq:wall} is explicitly treated by the finite difference approximation to the $z$ direction and Fourier spectral approximation to the $x$ and $y$ directions. 
The result shows $\Delta t_{max}=0.13$ and $\Delta t^{explicit}_{max}=0.003$, indicating the numerical scheme allows a larger time step size than the explicit scheme.

\subsection{Phase diagram of the PbTiO$_3$ thin film}
\begin{figure}[htbp]
\centering
\includegraphics[scale=0.79]{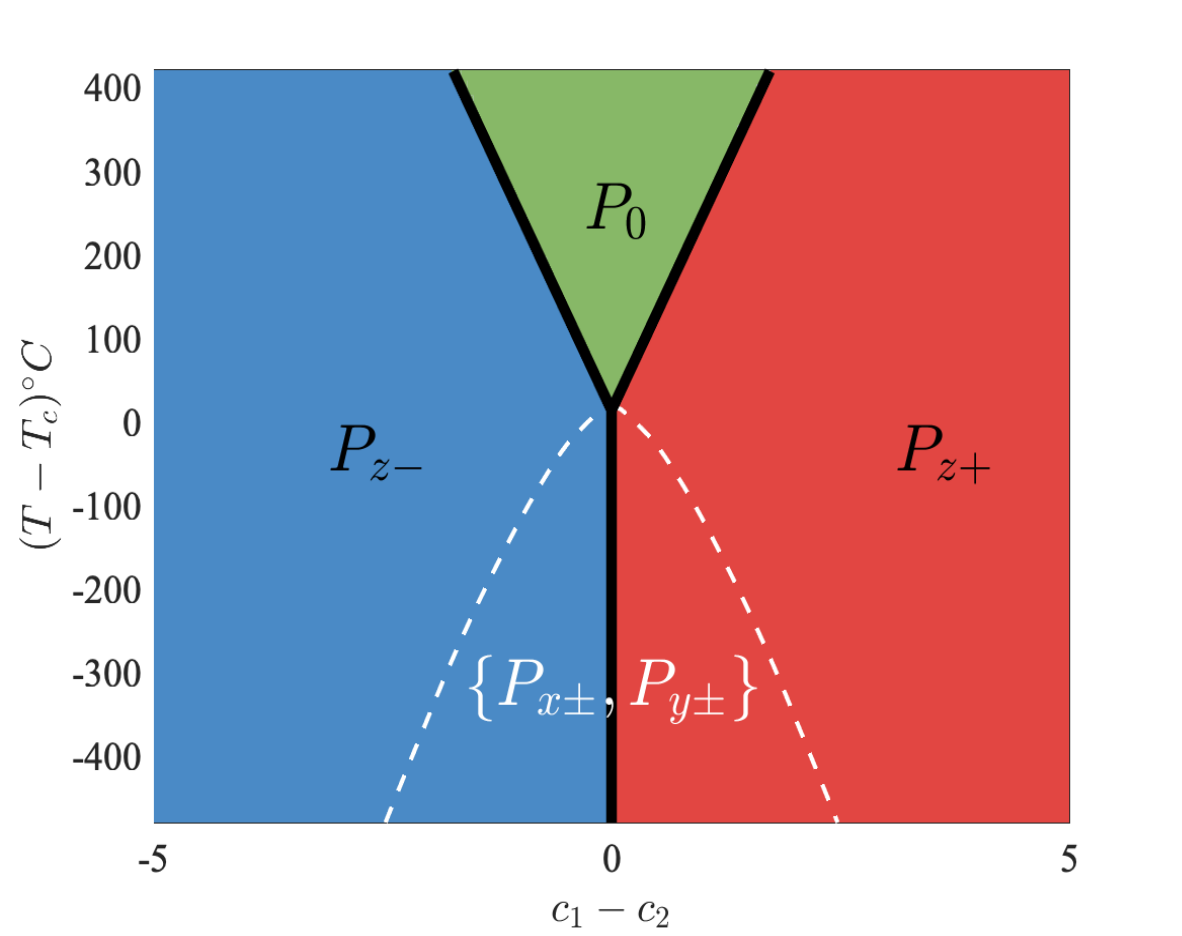}
\caption{\textbf{Phase diagram of PbTiO$_3$ thin film as a function of temperature and electric field with constant BC}. The Y-axis represents the difference between temperature ($T$) and Curie temperature ($T_c$), and the X-axis is the voltage between the bottom ($c_1$) and top ($c_2$) surfaces under constant BC. Each phase labeled by the notation $P_{z+}$, $P_{z-}$, and $P_0$ represents the polarization along the [001] direction, opposite to the [001] direction, and the paraelectric phase, respectively. The notation $P_{x\pm}$ and $P_{y\pm}$ represent the polarization along or opposite to the [100] and [010] directions, respectively. } 
\label{fig:3}
\end{figure}

To study the electric field effect on the stability of ferroelectric domain structures, a phase diagram is constructed in Fig. \ref{fig:3}. The voltage between the bottom ($c_1$) and the top ($c_2$) surfaces ranging from $-5$ to $5$ under constant BC, and the temperature ($T$) ranging from 0$^\circ C$ to 900$^\circ C$  are considered to investigate their effects on the domain configuration. It shows stable paraelectric or ferroelectric phases and their corresponding domain structures as a function of temperature and the voltage between the bottom and top surfaces under constant BC. 
Based on the Curie$-$Weiss law, the coefficient $\alpha_1$ in Eq. \eqref{eq:landau} has a linear temperature ($T$) dependence: $\alpha_1=\alpha(T-T_c)$, where $\alpha $ is related to Curie constant, and $T_c$ is the Curie$-$Weiss temperature, beyond which the spontaneous polarization of ferroelectric disappears and becomes a paraelectric phase ($P_0=(0,0,0)$). Here, $\alpha$ and $T_c$ are taken to be $3.8\times10^5C^{-2}m^2N/^{\circ}C$ and 479$^{\circ}$C, respectively \cite{Li_Actamateria_2002_subs}. 


Under a relatively large electric field, the phase $P_{z+}$ or $P_{z-}$, representing the ferroelectric domain with the polarization along the [001] direction (red color) or opposite to the [001] direction (blue color), is the only stable state, respectively. 
For a relatively small electric field and lower temperature, the area within white dashed line represents that the phases $P_{x\pm}$ and $P_{y\pm}$ with polarization along or opposite to the [100] and [010] directions respectively are the metastable states that have higher energies than $P_{z+}$ or $P_{z-}$.
On the other hand, the phase in green color shows that the paraelectric phase $P_0$ is the only stable state due to the high temperature effect. 
The results demonstrate that the stable polarization domain has to be formed to minimize the electric effect under a relatively large electric field, i.e., the polarization direction of the stable ferroelectric domain should be consistent with the electric field direction.

\subsection{Polarization switching pathways of the PbTiO$_3$ thin films}
Next, we compute the polarization switching pathways of ferroelectric thin films and the corresponding transition states. The transition state, which represents the critical nucleus, is the point (state) with the highest energy along the minimum energy path (MEP) that connects two steady states \cite{E_PhyRB_2002_string}.
Many numerical methods have been developed to compute the MEP and the transition state, including the Nudged Elastic Band method \cite{Henkelman_JChemP_2000_a} and the dimer-type methods \cite{henkelman1999dimer, weinan2011gentlest, zhang2012shrinking, Zhang_SIAMJSCom_2016_opt, Yin_SISC_2019, yu2020global}, etc.

In this paper, we adopt the string method in \cite{E_JChemP_2007_simplify} and combine it with the numerical scheme for the variational phase field model to study the polarization switching pathways under different electric BCs. 
The string method as well as its improvements has been developed for finding the MEPs and successfully applied to many applications \cite{E_PhyRB_2002_string, Du_ComMatSci_2009_a, ren2013climbing, Zhang_PRL_2007_morph, han2019transition, han2020pathways}.
We apply the string method to show the complete processes of $180^{\circ}$ polarization switching from $\bP=(0,0,-1)$ to $\bP=(0,0,1)$ and $90^\circ$ polarization switching from $\bP=(0,0,1)$ to $\bP=(0,1,0)$.

\begin{figure}[htbp]
\centering
\includegraphics[scale=0.7]{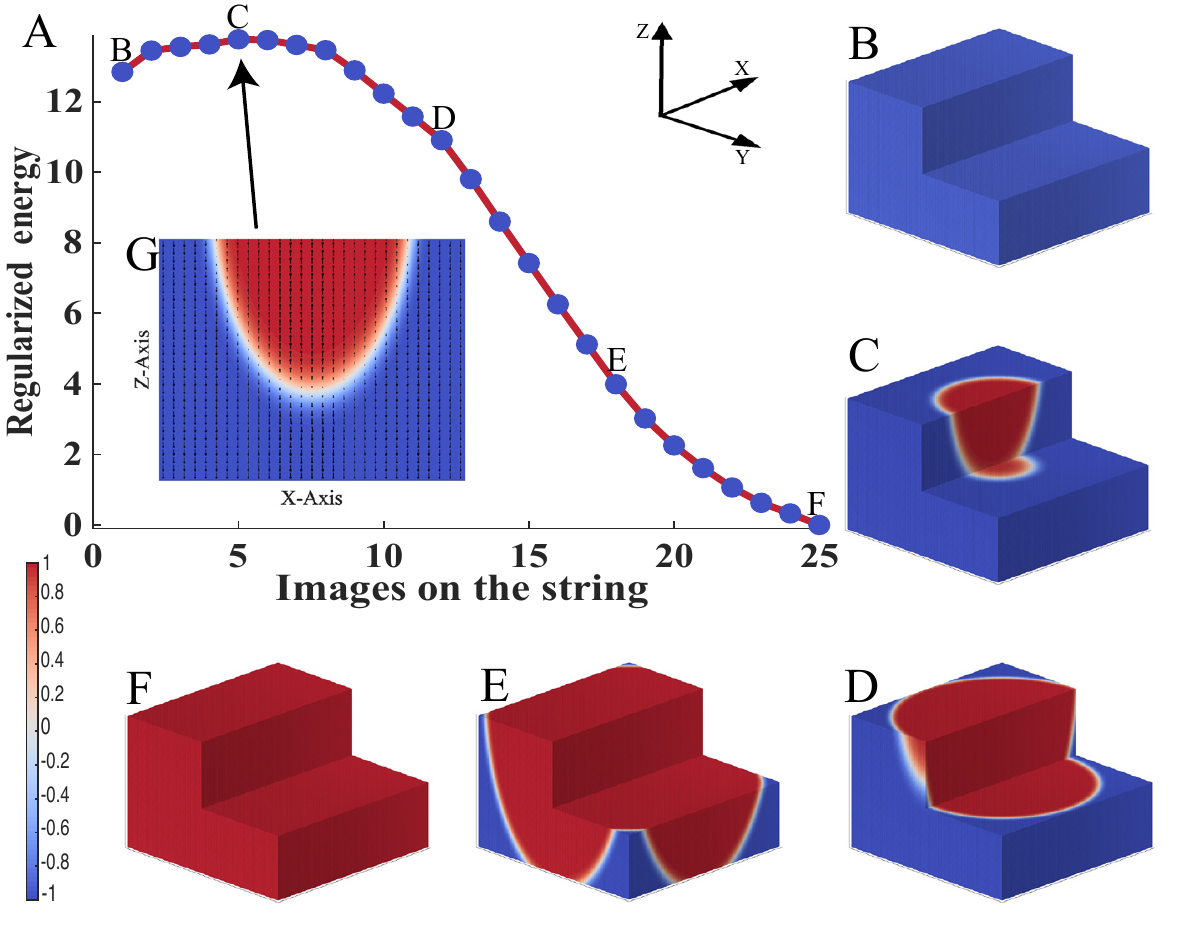}
\caption{\textbf{Computed MEP shows a boundary nucleation process of $180^\circ$ polarization switching under the tip-induced BC.}  (A) is the MEP of the polarization switching process connecting the initial polarization state ($\bP=(0,0,-1)$) (B) with the final polarization state ($\bP=(0,0,1)$) (F) by passing through a tip-induced critical nucleus (C) and two intermediate states (D) and (E). (G) is the sliced view of the critical nucleus (C) at $y=0$.}
\label{fig:4}
\end{figure}

\begin{figure}[htbp]
\centering
\includegraphics[scale=0.7]{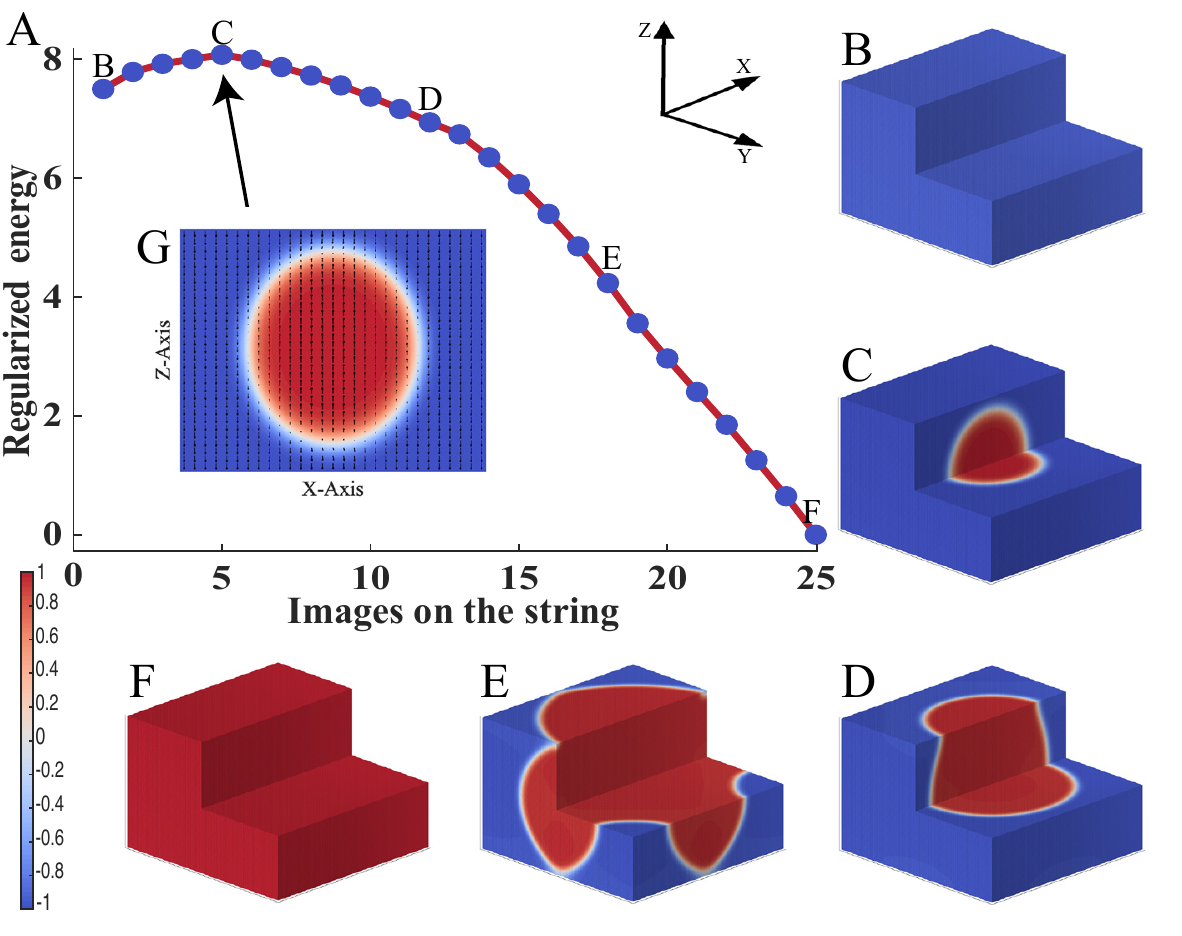}
\caption{\textbf{Computed MEP shows an inner nucleation process of the $180^\circ$ polarization switching under the constant BC.} (A) is the MEP of the polarization switching process connecting the initial polarization state ($\bP=(0,0,-1)$) (B) with the final polarization state ($\bP=(0,0,1)$) (F) by passing through an ellipsoid-shaped critical nucleus (C) and two intermediate states (D) and (E). (G) is the sliced view of the critical nucleus (C) at $y=0$.}
\label{fig:5}
\end{figure}

In Fig. \ref{fig:4} and Fig. \ref{fig:5}, we plot the MEPs of the $180^\circ$ polarization switching pathways, corresponding to the tip-induced nucleation and the inner ellipsoid-shaped nucleation, respectively. Fig. \ref{fig:4} shows the critical nucleus could be formed on the boundary owing to the tip-induced electric BC and becomes a half-ellipsoid nucleus. 
Meanwhile, if the constant BC is applied, Fig. \ref{fig:5} shows that the critical nucleus has an ellipsoidal  shape with their long axis along the direction of the electric field, i.e., the Z-axis. We slice the configurations of the critical nuclei along the $y=0$ axis to show its domain pattern in the 2D X-Z plane in Fig. \ref{fig:4}G and Fig. \ref{fig:5}G. The scales and directions of black arrows indicate the local polarization magnitudes and directions of each unit cell in the X-Z plane. Once a critical nucleus is formed to overcome the energy barrier, the switched polarization domain continues to grow until the final steady state is reached. \par

\begin{figure}[htbp]
\centering
\includegraphics[scale=0.7]{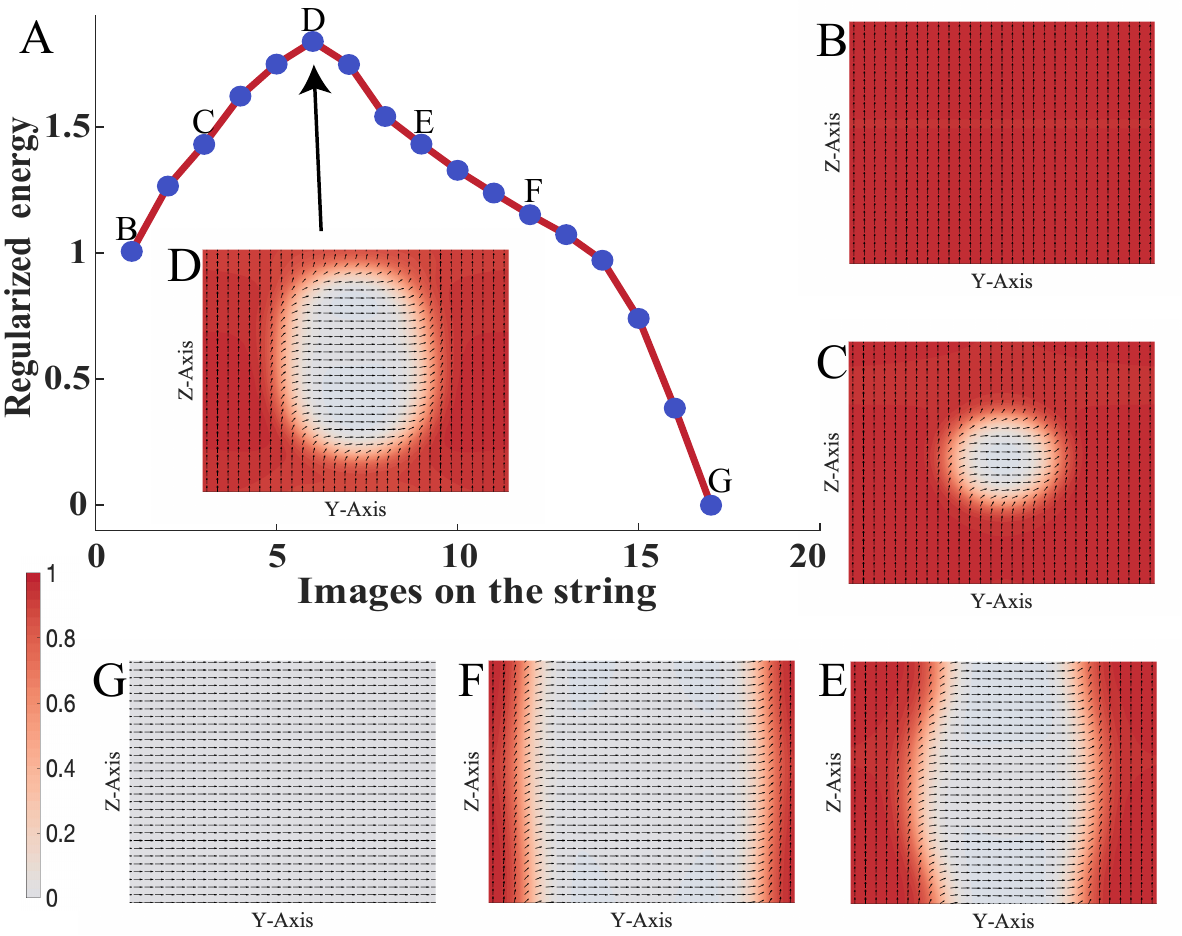}
\caption{\textbf{Computed MEP shows a nucleation process of $90^\circ$ polarization switching under the open circuit BC.} 
(A) is the MEP of the polarization switching process connecting the initial polarization state ($\bP=(0,0,1)$) (B) with the final polarization state ($\bP=(0,1,0)$) (G) via the sliced view at $x=0$.
(D) is the sliced view of an ellipsoid-shaped critical nucleus and (C,E,F) are the sliced views of the intermediate states corresponding to points C, E, and F on the MEP, respectively.}
\label{fig:6}
\end{figure}

In Fig. \ref{fig:6}, we plot the MEP of the $90^{\circ}$ polarization switching pathway starting from the domain with polarization along the Z-axis to the domain with polarization along the Y-axis in the presence of the electric field. 
We use the sliced view at $x=0$ to show its domain pattern in the Y-Z plane.
Red and grey colors are used to represent the equivalent polarization magnitude and the corresponding polarization direction, i.e., $\bP=(0,0,1)$ and $\bP=(0,1,0)$, respectively. 
Fig. \ref{fig:6} shows that the nucleus is first formed in an ellipsoidal shape with polarization along the Y-axis (Fig. \ref{fig:6}C), and then expands with its long axis being arranged along the direction of the electric field, i.e., the Z-axis. Once the critical nucleus is formed (Fig. \ref{fig:6}D), the switched polarization domain continues to grow and gradually changes to a stripe pattern, and finally reaches the final steady state $\bP=(0,1,0)$.\par

\section{Conclusions}
\label{sec:conclus}
In this paper, we present a new, efficient numerical scheme for the variational phase field model based on the variational phase field formulations of ferroelectric thin films. It avoids solving the electrostatic equilibrium equation (a Poisson equation) and the use of associated Lagrange multipliers during numerical iterations. Moreover, by making use of the explicitly formulated expressions of the driven forces, all the calculations actually become the multiplication of matrices and vectors at the discrete level. The numerical scheme for the relaxation dynamics of the polarization vector is designed in a semi-implicit way, which allows us to use a larger time step size at each iteration compared to the explicit scheme. 

The proposed numerical scheme is then applied to study phase transitions and polarization switching processes under different electric BCs. Numerical results show that the electric field on the ferroelectric thin film can generate a local tip-like polarization domain under the tip-induced BC, or a ribbon-like polarization domain under the constant BC. 
We demonstrate that the traditional way of taking electric potential $\phi$ and polarization vector $\bP$ as independent variables could underestimate the effect of the electric field on the ferroelectric phase transitions, as the electrostatic driven force $e_{old}(\bP)$ is only half of the accurate one $e(\bP)$.
We also show two types of the $180^\circ$ polarization switching pathways, i.e., boundary nucleation and inner nucleation processes. Furthermore, from the example of the $90^{\circ}$ polarization switching pathway, we find that the polarization along the Y-axis (or equivalently the X-axis) is more stable than that along the Z-axis under the open circuit BC, which is consistent with the numerical observations made in \cite{Li_APL_2002_effect}. 

The current numerical discretization of Eq. \eqref{eq:f1} adopts the semi-implicit scheme for time integration. We mostly focused on the effective spatial discretization. It will be interesting to also explore more sophisticated time discretization schemes \cite{Qiao_2011_Siamsci} that preserve the energy stability for the generalized energy system \eqref{eq:energy}. 
The current version of variational phase field formulations only focus on the electrostatic field in ferroelectric thin films. Other energy contributions, e.g., the elastic energy and the magnetic energy, could also play a crucial role in ferroelectric phase transitions. For example, the electromechanical coupling between the electric field and the elastic strain offers a powerful route for the selective control of multiple domain switching pathways in multiaxial ferroelectric materials \cite{flexoelectric_nanotech_2018}. The magnetic polar-skyrmions could contribute to the advancement of ferroelectrics towards functionalities \cite{skyrmion_nature_2019}. Thus,  a systematic variational phase field approach that can deal with various energy contributions and construction of a solution landscape \cite{Yin_prl_2020, yin2020searching} for ferroelectric thin films will be interesting to pursue in future.\par

{\bf Acknowledgment}
We would like to thank Prof. Long-Qing Chen for bringing the subject studied here to our attention. Thanks to Dr. Yu-Lan Li and Dr. Bo Wang for fruitful discussions. This work of Lei Zhang was supported by the National Natural Science Foundation of China No. 11861130351 and the Royal Society Newton Advanced Fellowship. The work of Qiang Du is supported in part by NSF DMS-1719699 and NSF CCF-1704833. Ruotai Li also acknowledges the funding support from the China Scholarship Council No.201806010041.

\bibliographystyle{unsrt}
\bibliography{references_cicp}

\end{document}